\documentclass[aip,pop,reprint]{revtex4-1}
\usepackage{graphicx}
\usepackage{subfigure}
\usepackage{amsmath}
\usepackage{overpic}
\usepackage{color}
\usepackage{rotating}
\usepackage{txfonts}
\usepackage{array}
\usepackage{multirow}
\usepackage{url}
\usepackage[colorlinks=true, allcolors=blue]{hyperref}
\draft 

\begin{document}

\title{High Flux Electron Beams from Laser Wakefield Accelerators Driven by Petawatt Lasers} 

\author{Ming Zeng}
\email{ming.zeng@eli-np.ro}
\affiliation{Extreme Light Infrastructure - Nuclear Physics, Horia Hulubei National Institute for Physics and Nuclear Engineering, 30 Reactorului Street, P.O. Box MG-6, 077125 Magurele, jud. Ilfov, Romania}
\author{Ovidiu Tesileanu}
\affiliation{Extreme Light Infrastructure - Nuclear Physics, Horia Hulubei National Institute for Physics and Nuclear Engineering, 30 Reactorului Street, P.O. Box MG-6, 077125 Magurele, jud. Ilfov, Romania}

\begin{abstract}
Laser Wakefield Accelerator (LWFA) is considered as one of the most competitive candidates for the accelerators of the next generation. With the development of high power laser technologies, LWFA has shown its potential of replacing the conventional radio-frequency (RF) accelerators due to its flexibility and adjustability. In this paper, we will study the potential high flux electron beam productions of LWFA driven by petawatt-level laser pulses. In our three dimensional particle-in-cell simulations, an optimal set of parameters gives $\sim 40\ \rm nC$ of charge with $2\ \rm PW$ laser power, thus $\sim 400\ \rm kA$ of instantaneous current if we assume the electron beam duration is 100 fs. This high flux and its secondary radiation are widely applicable in nuclear and QED physics, industrial imaging, medical and biological studies.
\end{abstract}

\maketitle
\section{Introduction} 

Ever since its invention, the laser wakefield accelerator (LWFA) has been regarded as one of the best candidates for the accelerators of the next generation~\cite{TajimaPRL1979}. Compared with the conventional radio frequency (RF) accelerators, the LWFA has the advantages of much larger acceleration gradient and thus more compact size if the same output energy is required~\cite{EsareyRMP2009}. Starting from 2004 and thanks to the development of the high-power femtosecond laser technologies, the output beam quality of the LWFA is approaching that of the RF accelerator~\cite{ManglesNature2004,GeddesNature2004,FaureNature2004,HTKimPRL2013,LeemansPRL2014}, making it widely interesting for applications in a variety of fields. Apart from the high energy (GeV level) and low energy spread ($\sim 1\%$) requirements of some applications, large electron flux is one of the most widely demanded characters of the accelerator output beams~\cite{KHommaRRP2016}. Especially for nuclear or QED related researches, large electron flux can greatly increase the yielding of the products within a certain time duration, while neither the high electron beam energy nor the low energy spread is required. It has been reported that about 20 nC charge electron beams can be produced by 100 TW level lasers~\cite{BFShenPOP2012}. However, this report was based on the transverse profile complexity of the laser beams. For simple Gaussian mode lasers, the dependence of the output charge number with the laser power and plasma density has not been studied comprehensively yet.

In this paper, we study the potential of generating high flux electron beam by petawatt-level laser pulses in the LWFAs using the three-dimensional (3D) particle-in-cell (PIC) simulation method. It is found that there is a threshold peak power $P_{\rm th}$, that with laser peak power $P > P_{\rm th}$ the laser energy can be transversely well confined in the plasma, while for $P < P_{\rm th}$ the laser spot size increases continuously. $P_{\rm th}$ is negatively correlated with the plasma density $n_p$ and is similar to the relativistic self-focusing power~\cite{SprangleIEEE1987,GZSunPOF1987,KCTzengPRL1998,MZengPOP2014}. However, in the current paper $P_{\rm th}$ is obtained from 3D PIC simulations which consider the influences of both effects of relativistic mass increase and plasma cavitation. In addition, the curve of the charge number of the output electron beam vs.\ laser peak power is found to have a transition at $P_{\rm th}$. This gives us an optimal choice of laser peak power for the maximum average current output.

   \begin{figure*}
   \begin{center}
   \begin{overpic}[width=0.8\textwidth]{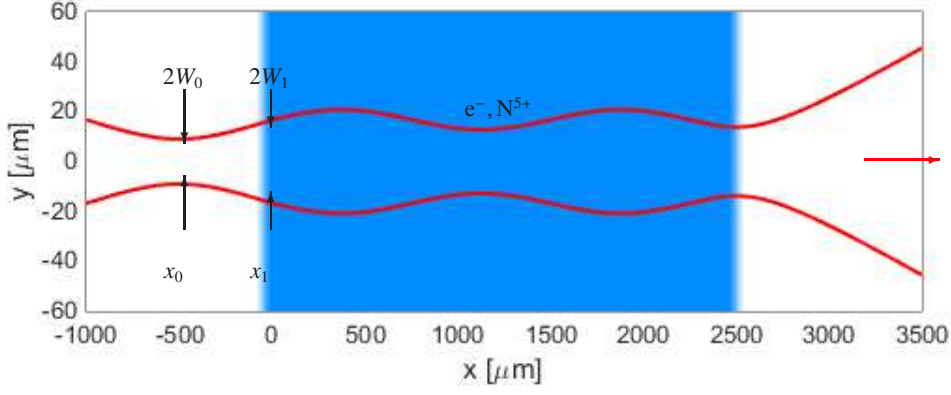}
      \put(22,14.5){\vector(0,1){5}}
      \put(22,27.5){\vector(0,-1){5}}
      \put(30,14.5){\vector(0,1){3.5}}
      \put(30,27.5){\vector(0,-1){3.5}}
      \put(85,21){\color{red}\vector(1,0){7}}
      \put(20,28){$2W_0$}
      \put(28,28){$2W_1$}
      \put(20,10){$x_0$}
      \put(28,10){$x_1$}
      \put(48,25){$\rm e^-, N^{5+}$}
   \end{overpic}
   \end{center}
   \caption[Waveform]
   { \label{fig:schematic} Schematic view of the laser wakefield accelerator with the laser focus located in front of the plasma region. The plasma density has a flat-top from $x = 0$ to $2450\ \rm \mu m$, and for $x < -100\ \rm \mu m$ or $x > 2550\ \rm \mu m$ there is vacuum. Between the vacuum and the plasma density flat-top there are density transition regions with the $\sin^2$ profile. The laser has a simple Gaussian mode with waist of $W_0$ and is focused at $x_0 < -100\ \rm \mu m$ in the vacuum region on the left, and when it arrives $x_1 = 0\ \rm \mu m$ its spot size increases to $W_1$. During its propagation in the plasma region, the laser can be confined due to the self-focusing effect, and excites wakefield to accelerate electron beams (not shown in the figure). After exiting the plasma region, the laser diverges and the electron beams come out with high energies. The plasma electrons are ionized from pure nitrogen gas up to 5+ charge state by the laser pre-pulses, and the remaining K-shell electrons of $\rm N^{5+}$ can be captured by the wake via the ionization injection process. }
   \end{figure*}

\section{Analytical discussions}
\label{sec:sf}

In current technologies high power laser beams can reach the peak power in the petawatt level. If we assume a perfect Gaussian profile for the laser beam, it has the relation $P \left[\rm GW\right] = 21.5 \left( a W / \lambda \right)^2$, where $P$ is the peak power, $a$ is the normalized peak laser vector potential, $W$ is the spot size factor of a Gaussian beam, and $\lambda$ is the wavelength~\cite{EsareyRMP2009}. For the large charge number production purpose, we need a relatively large laser spot size~\cite{MZengJPCS2016}. And to drive a highly relativistic wake, $1.5 < a < 2.5$ is enough and good for ionization injections~\cite{MinChenJAP2006,McGuffeyPRL2010,PakPRL2010,ClaytonPRL2010,JSLiuPRL2011,PollockPRL2011,MZengJPP2012,MChenJCP2013,MZengPOP2014,MZengPRL2015,MZengPOP2016}. Thus $W$ can be $\sim 100\ \rm \mu m$ for a $800\ \rm nm$ wavelength petawatt laser beam. Meanwhile to replace the laser focusing system is difficult and expensive in experiments, if we want to change the laser power but keep $a$ unchanged. The solution is that we use a fixed focusing system to focus the laser to a smaller waist size $W_0 < W$, and place the plasma entrance a distance after the focal spot so that when the laser reaches the plasma, it has $a\approx 2$.

Our configuration is schematically shown in Fig.~\ref{fig:schematic}. The blue region shows the plasma with density of $n_p$ which is either pre-ionized or ionized by laser pre-pulse from nitrogen gas, while the white region is vacuum. There are transition regions between the vacuum and the plasma at both the front and rear sides. The high power laser beam is focused at some place before the plasma region ($x_0<0$) with the focal waist $W_0$, so that when the laser reaches $x_1=0$ the spot size factor $W_1 > W_0$ and the normalized laser vector potential $a_1 = 2$. With simple calculations one may find that $W_1 = \sqrt{P \left[\rm GW\right]/21.5}\times \lambda / a_1$, $x_0=x_1-x_R\sqrt{\left(W_1/W_0\right)^2-1}$, where $x_R=\pi W_0^2/\lambda$ is the Rayleigh length.

In weakly relativistic cases ($a \ll 1$ and thus the plasma density response is negligible), the laser profile evolves according to the equation~\cite{SprangleIEEE1987}
\begin{equation}\label{eq:W}
  \frac{\rm d^2}{{\rm d} x^2}{W} = \frac{\lambda^2}{\pi^2 W^3} \left[ 1 - \frac{\omega_p^2a^2W^2}{32c^2} \right] = \frac{\lambda^2}{\pi^2 W^3} \left[ 1 - \frac{P}{P_c} \right],
\end{equation}
where $P_c \equiv \frac{8\pi m_e^2c^3}{e^2\mu_0} \frac{\omega^2}{\omega_{\rm p}^2} \approx 17.4 \frac{\omega^2}{\omega_p^2} \left[ \rm GW \right]$ is the relativistic self-focusing critical power of a laser beam in plasma, $\omega$ is the laser frequency and $\omega_p$ is the plasma frequency. If we assume $P/P_c$ is a constant during the propagating process, i.~e.\ the plasma density is a constant and the energy loss of the laser beam is negligible, the evolution of the laser spot size can be solved with certain initial conditions. For example, if $W=W_0$ and $\frac{\rm d}{{\rm d} x}W=0$ at $x=0$, $W$ is monotonically increasing for $x>0$ in the case $P<P_c$, and $W$ is monotonically decreasing for $x>0$ in the case $P>P_c$. This initial condition corresponds to the situation that the laser is focused at the vacuum-plasma boundary, and $W_0$ is the laser beam waist. However, if we focus the laser before the vacuum-plasma boundary, the initial condition at $x=0$ becomes $W=W_1>W_0$ and $\frac{\rm d}{{\rm d} x}W>0$. For the case $P<P_c$, $W$ is still monotonically increasing, while for the case $P>P_c$, $W$ increases to its maximum $W_{\max}>W_1$ and starts to decrease afterwards.

With the above discussions, we may conclude that by focusing the laser beam before the vacuum-plasma boundary with a waist of $W_0$, one can achieve the same effect as focusing the laser beam at the vacuum-plasma boundary with a waist $W_{\max}$, asserting $P>P_c$. This is a very useful method for achieving a suitable effective laser spot size without replacing the focusing system in real LWFA facilities.

  \begin{figure*}
    \begin{center}
      \begin{tabular}{cc}
        \subfigure{\label{fig:Wnp2}}
        \begin{overpic}[width=0.5\textwidth]{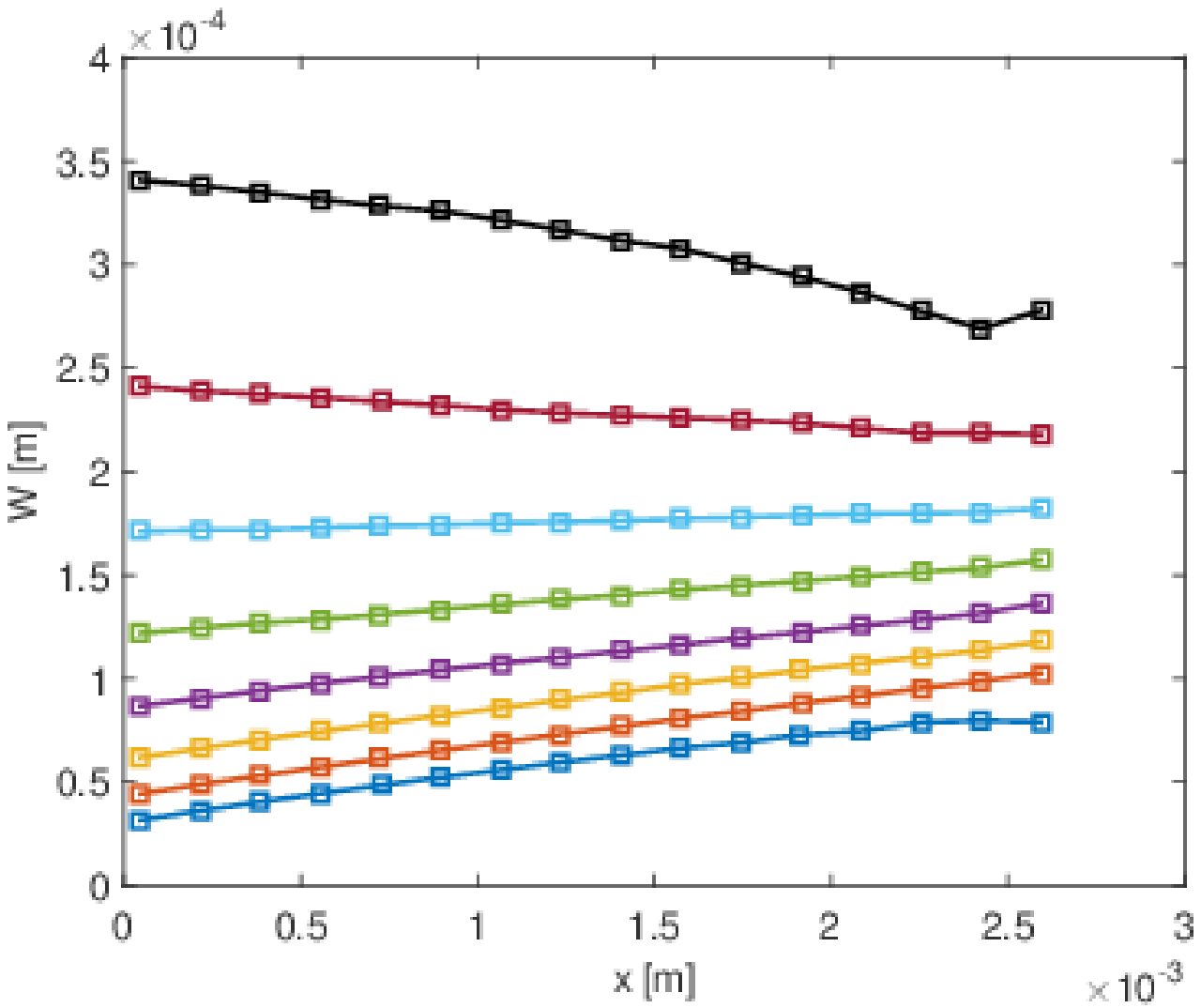}
          \put(78,60){(a)}
        \end{overpic}
        \subfigure{\label{fig:Wnp4}}
        \begin{overpic}[width=0.5\textwidth]{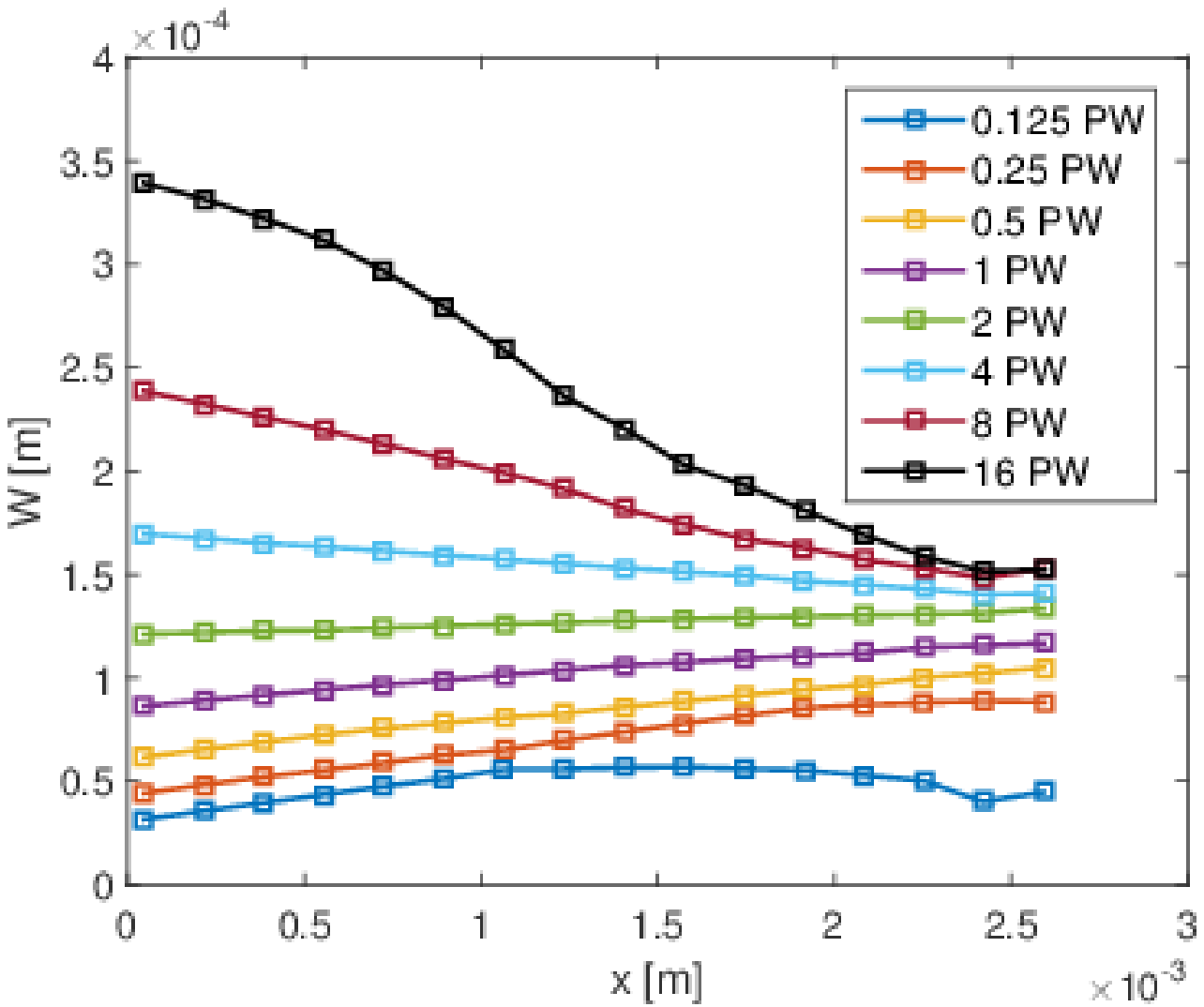}
          \put(22,60){(b)}
        \end{overpic}
      \end{tabular}
    \end{center}
    \caption[Waveform]
      { \label{fig:W_evol} Laser spot size parameter $W$ evolution with different laser peak power and plasma density cases. (a) $n_p = 2 \times 10^{18}\ \rm cm^{-3}$ and laser peak power varies from $0.125$ to $16\ \rm PW$. (b) $n_p = 4 \times 10^{18}\ \rm cm^{-3}$ and laser peak power varies from $0.125$ to $16\ \rm PW$. Data obtained from 3D PIC simulations }
  \end{figure*}

\section{Simulations}

Section \ref{sec:sf} has discussed the laser profile evolution if the laser is focused before entering the plasma region using the weakly relativistic model. However, the actual laser profile function can be different from Eq.~(\ref{eq:W}) especially in the situation $a\gtrsim 1$, i.~e.\ the density modification is not negligible. Especially when $a\gtrsim 2$, the laser beam blows the plasma electrons out of the central region and leaves a electron-vacant region, but the ions are still almost stationary because of their much smaller charge-mass ratio compared with electrons. This is called the blowout regime or ``bubble'' regime. In this regime, the actual self-focusing power is much higher than that in the weakly relativistic model, and is more difficult to calculate analytically. We performed 3D PIC simulations with the code EPOCH~\cite{EPOCH} to study the laser spot size evolution in this regime. We chose two plasma densities, and the results are shown in Fig.~\ref{fig:W_evol}. The simulation box has the longitudinal dimension of $70\ \rm \mu m$, and the transverse dimension of $6W_1\times 6W_1$ (varies from case to case, so that the laser beam can be well limited in the simulation box, and the transverse resolution is enough for the spot size on the same time). The simulation resolution is fixed to $1024\times 128 \times 128$, and simulation time step $\Delta t$ is set to be very close but a little bit smaller than the Courant Condition requirement. In the plasma region, the number of macro particle per cell is 4 for the plasma electrons and $\rm N^{5+}$, and the background positive charge automatically neutralizes the total charge due to the simulation algorithm. The density of $\rm N^{5+}$ is $n_{\rm N^{5+}}=n_p / 5$, where $n_p$ is the plasma density, thus all the plasma electrons are ionized from the pure nitrogen gas by the laser pre-pulse. The plasma profile is schematically shown in Fig.~\ref{fig:schematic}, and the laser is focused at $x=x_0<0$, so that when it reaches $x=x_1=0$, the normalized vector potential is $a=a_1=2$.

Figures~\ref{fig:Wnp2} and \ref{fig:Wnp4} show the cases with $n_p = 2 \times 10^{18}\ \rm cm^{-3}$ and $4 \times 10^{18}\ \rm cm^{-3}$, respectively. One should notice that the plasma only exists in the region $-0.1\ {\rm mm} < x < 2.55\ {\rm mm}$, thus the right most dot of each curve is actually outside the plasma region. One can see that the threshold power $P_{\rm th}$ for the good confinement of the laser beam (i.~e.\ $W$ be monotonically decreasing in the plasma region) is (a) $4\ {\rm PW} \lesssim P_{\rm th} < 8\ {\rm PW}$ for $n_p = 2 \times 10^{18}\ \rm cm^{-3}$ and (b) $2\ {\rm PW} \lesssim P_{\rm th} < 4\ {\rm PW}$ for $n_p = 4 \times 10^{18}\ \rm cm^{-3}$. For $P>P_{\rm th}$, the laser spot size changes from $W_0$ to $W_1$ in the vacuum region, and then changes from $W_1$ to $W_{\max}$ in a very limited region after entering plasma that this process is hardly observable in these plots. We cannot observe the oscillation of $W$ because the plasma region length is smaller than the oscillation period.

   \begin{figure*}
   \begin{center}
      \begin{tabular}{cc}
        \subfigure{\label{fig:power_charge}}
        \begin{overpic}[width=0.5\textwidth]{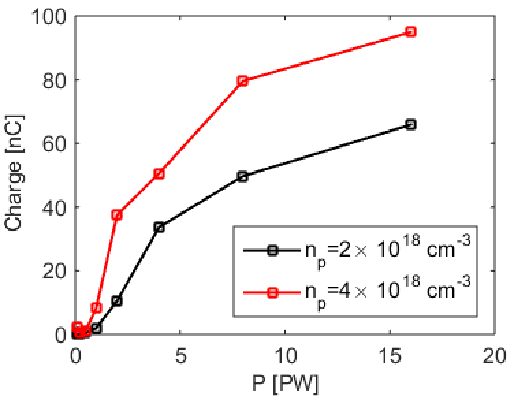}
          \put(78,60){(a)}
        \end{overpic} &
        \subfigure{\label{fig:spectrum}}
        \begin{overpic}[width=0.5\textwidth]{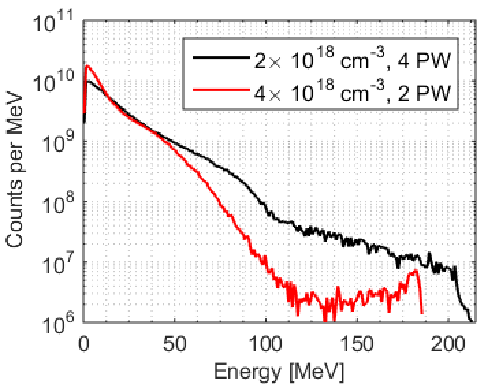}
          \put(22,60){(b)}
        \end{overpic} \\
        \subfigure{\label{fig:ne1}}
        \begin{overpic}[width=0.5\textwidth]{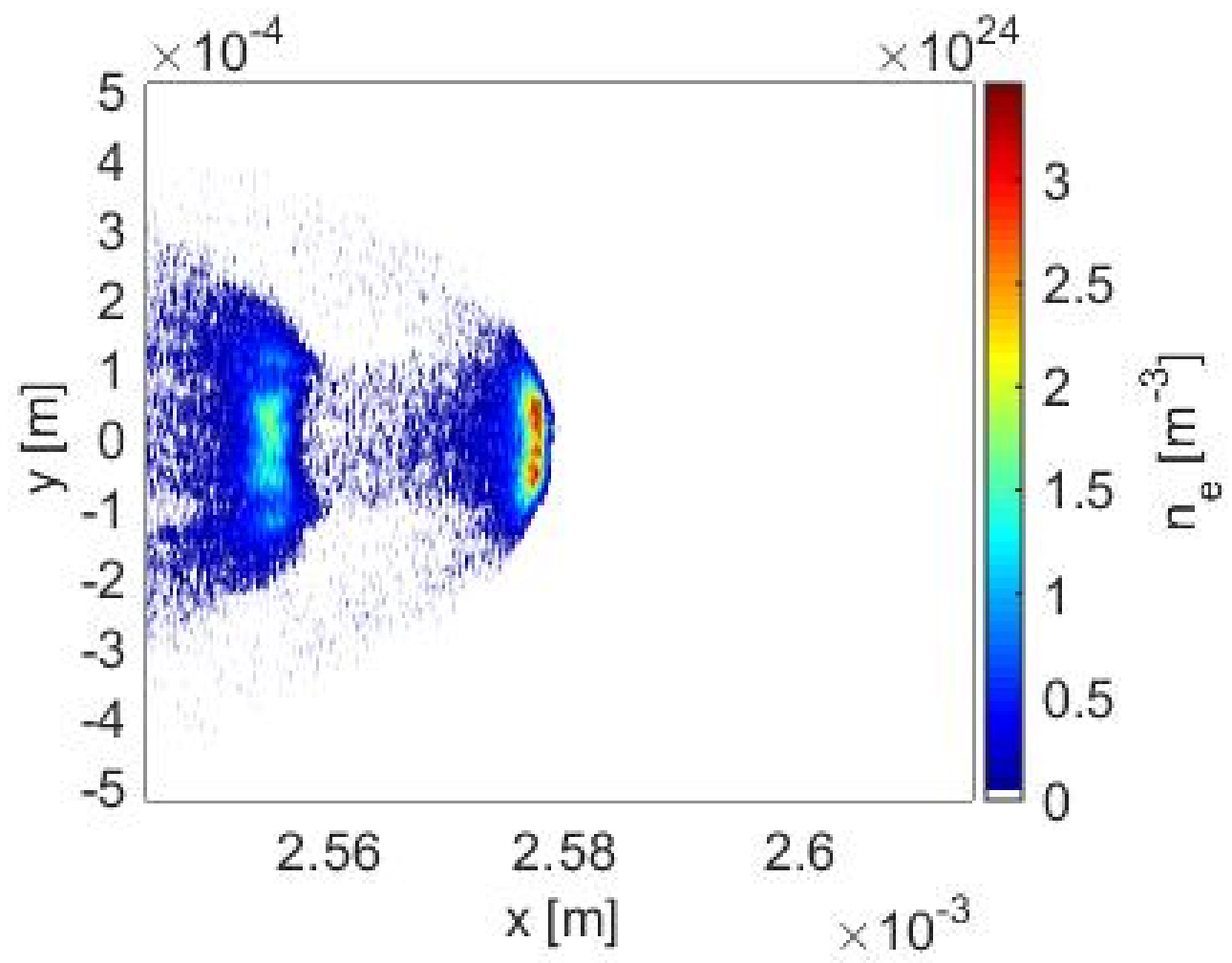}
          \put(68,60){(c)}
        \end{overpic} &
        \subfigure{\label{fig:ne2}}
        \begin{overpic}[width=0.5\textwidth]{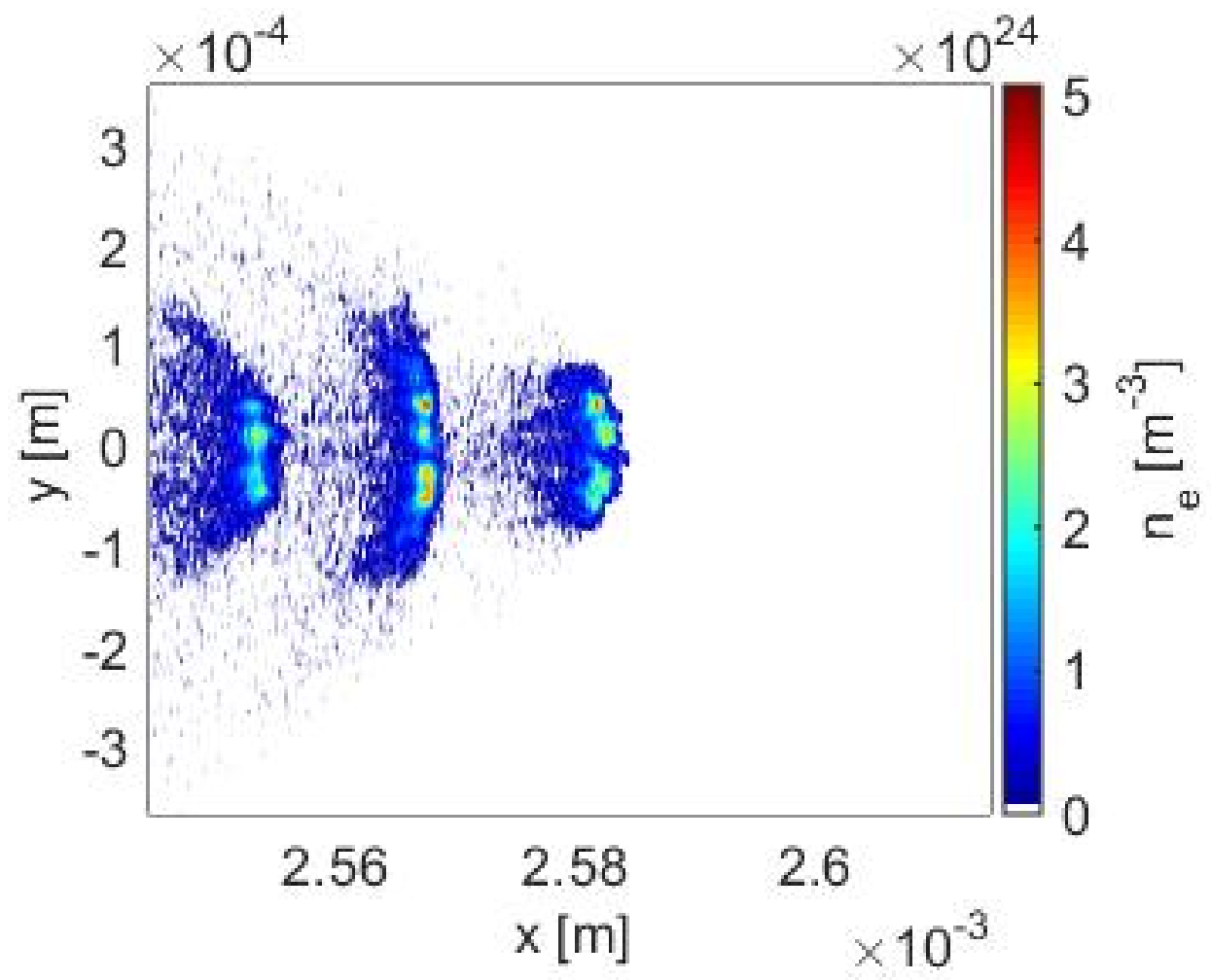}
          \put(70,60){(d)}
        \end{overpic}
      \end{tabular}
   \end{center}
   \caption[Waveform]
   { \label{fig:ebeam} Output electron beam properties. (a) Charge number vs.\ laser peak power. The black and red lines are for $n_p = 2 \times 10^{18}\ \rm cm^{-3}$ and $n_p = 4 \times 10^{18}\ \rm cm^{-3}$ cases, respectively. (b) Electron energy spectra for the cases with $n_p = 2\times 10^{18}\ \rm cm^{-3}$, $P=4\ \rm PW$ (black line) and $n_p = 4\times 10^{18}\ \rm cm^{-3}$, $P=2\ \rm PW$ (red line). (c) and (d) are the output electron beam distribution slices at $z=0$ for the cases $n_p = 2\times 10^{18}\ \rm cm^{-3}$, $P=4\ \rm PW$ and $n_p = 4\times 10^{18}\ \rm cm^{-3}$, $P=2\ \rm PW$, respectively. }
   \end{figure*}

Then we study the charge number of the output electron beams in all the above cases. Figure~\ref{fig:power_charge} shows the output electron beam charge numbers with the change of $n_p$ and $P$. One may see that for $n_p = 2 \times 10^{18}\ \rm cm^{-3}$, charge varies approximately proportional to $P^m$ ($m>0$) with the transition at $P=P_{\rm th}\approx 2\ \rm PW$, where for $P<P_{\rm th}$ cases $m>1$ and for $P>P_{\rm th}$ cases $m<1$. Meanwhile for $n_p = 4 \times 10^{18}\ \rm cm^{-3}$ the transition is at $P=P_{\rm th}\approx 4\ \rm PW$. In the current discussion $P_{\rm th}$ is the same as the power threshold in the discussion of Fig.~\ref{fig:W_evol}.

We assume that in a certain femtosecond laser facility, the average laser power is a constant, i.~e.\ $P_{\rm avg}=P\tau f = \rm const.$, where $P$ is the peak laser power, $\tau$ is the FWHM time duration of one laser pulse, and $f$ is the laser repetition rate. The average output electron beam current becomes $I_{\rm avg}=qf=q P_{\rm avg}/P\tau \propto q/P$ if we also assume $\tau$ does not change with $P$, where $q$ is the output charge number within one shoot. Thus we conclude that the maximum $I_{\rm avg}$ is achieved when $q/P$ reaches its maximum. By looking at Fig.~\ref{fig:power_charge}, one can find that the $q/P$ reaches its maximum at the transition $P=P_{\rm th}$. Consequently, we conclude that the optimal peak power for maximum electron flux is $P=P_{\rm th}$. The reason can be explained as follows. When the laser peak power equals to the threshold power $P_{\rm th}$ for laser beam confinement, the laser transverse size almost does not change, so the charge accumulation section does not change~\cite{MZengJPCS2016}. This case has a stable charge accumulation rate and has the most efficient charge injection per unit laser power. If $P<P_{\rm th}$, the laser spot size increases during the propagation process and the laser peak amplitude decreases. When the laser peak amplitude drops below the ionization injection threshold ($a_{\rm th} \approx 1.8$ for 800 nm lasers), the charge accumulation stops. And If $P>P_{\rm th}$, the laser spot size decreases during the propagation process, thus the charge accumulation section is reduced, and consequently the charge accumulation rate is reduced.

The energy spectra of the output electron beams for the cases $n_p = 2\times 10^{18}\ \rm cm^{-3}$, $P=4\ \rm PW$ and $n_p = 4\times 10^{18}\ \rm cm^{-3}$, $P=2\ \rm PW$ are shown in Fig.~\ref{fig:spectrum}. We can see that for the red line case (lower power and higher density), more electrons are concentrated in the low energy region compared with the black line (higher power and lower density). This makes the red line case more advantageous for nuclear applications requiring $\sim 10\ \rm MeV$ level radiations~\cite{KHommaRRP2016}. Figure~\ref{fig:ne1} and \ref{fig:ne2} are the electron beam snapshots when the beams have just exited the plasma region. Figure~\ref{fig:ne1} corresponds to the black line case and \ref{fig:ne2} corresponds to the red line case in Fig.~\ref{fig:spectrum}. One can see a few bunches of electron beams in the snapshots. These bunches are from the different bubbles of the lase wakefield.

\section{Conclusions}
We have studied the charge number production of petawatt level LWFAs with the single Gaussian profile laser beams. We found that there is a threshold laser peak power $P_{\rm th}$ for laser beam confinement, and $P_{\rm th}$ is related to the plasma density. We also found that the cases with laser peak power equals to $P_{\rm th}$ have the most efficient charge injection, and thus can produce the optimal output beam flux. Approximately 10 to 100 nanocoulomb of charge can be produced by the LWFA with a single petawatt laser pulse.

\acknowledgments
This work is supported by Extreme Light Infrastructure - Nuclear Physics (ELI-NP) Phase II, a project co-financed by the Romanian Government and European Union through the European Regional Development Fund. The EPOCH code project was funded by the UK EPSRC grants EP/G054950/1, EP/G056803/1, EP/G055165/1 and EP/ M022463/1.

\bibliography{Cai_PST}

\begin{thebibliography}{25}%
\makeatletter
\providecommand \@ifxundefined [1]{%
 \@ifx{#1\undefined}
}%
\providecommand \@ifnum [1]{%
 \ifnum #1\expandafter \@firstoftwo
 \else \expandafter \@secondoftwo
 \fi
}%
\providecommand \@ifx [1]{%
 \ifx #1\expandafter \@firstoftwo
 \else \expandafter \@secondoftwo
 \fi
}%
\providecommand \natexlab [1]{#1}%
\providecommand \enquote  [1]{``#1''}%
\providecommand \bibnamefont  [1]{#1}%
\providecommand \bibfnamefont [1]{#1}%
\providecommand \citenamefont [1]{#1}%
\providecommand \href@noop [0]{\@secondoftwo}%
\providecommand \href [0]{\begingroup \@sanitize@url \@href}%
\providecommand \@href[1]{\@@startlink{#1}\@@href}%
\providecommand \@@href[1]{\endgroup#1\@@endlink}%
\providecommand \@sanitize@url [0]{\catcode `\\12\catcode `\$12\catcode
  `\&12\catcode `\#12\catcode `\^12\catcode `\_12\catcode `\%12\relax}%
\providecommand \@@startlink[1]{}%
\providecommand \@@endlink[0]{}%
\providecommand \url  [0]{\begingroup\@sanitize@url \@url }%
\providecommand \@url [1]{\endgroup\@href {#1}{\urlprefix }}%
\providecommand \urlprefix  [0]{URL }%
\providecommand \Eprint [0]{\href }%
\providecommand \doibase [0]{http://dx.doi.org/}%
\providecommand \selectlanguage [0]{\@gobble}%
\providecommand \bibinfo  [0]{\@secondoftwo}%
\providecommand \bibfield  [0]{\@secondoftwo}%
\providecommand \translation [1]{[#1]}%
\providecommand \BibitemOpen [0]{}%
\providecommand \bibitemStop [0]{}%
\providecommand \bibitemNoStop [0]{.\EOS\space}%
\providecommand \EOS [0]{\spacefactor3000\relax}%
\providecommand \BibitemShut  [1]{\csname bibitem#1\endcsname}%
\let\auto@bib@innerbib\@empty
\bibitem [{\citenamefont {Tajima}\ and\ \citenamefont
  {Dawson}(1979)}]{TajimaPRL1979}%
  \BibitemOpen
  \bibfield  {author} {\bibinfo {author} {\bibfnamefont {T.}~\bibnamefont
  {Tajima}}\ and\ \bibinfo {author} {\bibfnamefont {J.~M.}\ \bibnamefont
  {Dawson}},\ }\href {\doibase 10.1103/PhysRevLett.43.267} {\bibfield
  {journal} {\bibinfo  {journal} {Phys. Rev. Lett.}\ }\textbf {\bibinfo
  {volume} {43}},\ \bibinfo {pages} {267} (\bibinfo {year} {1979})}\BibitemShut
  {NoStop}%
\bibitem [{\citenamefont {Esarey}\ \emph {et~al.}(2009)\citenamefont {Esarey},
  \citenamefont {Schroeder},\ and\ \citenamefont {Leemans}}]{EsareyRMP2009}%
  \BibitemOpen
  \bibfield  {author} {\bibinfo {author} {\bibfnamefont {E.}~\bibnamefont
  {Esarey}}, \bibinfo {author} {\bibfnamefont {C.~B.}\ \bibnamefont
  {Schroeder}}, \ and\ \bibinfo {author} {\bibfnamefont {W.~P.}\ \bibnamefont
  {Leemans}},\ }\href {\doibase 10.1103/RevModPhys.81.1229} {\bibfield
  {journal} {\bibinfo  {journal} {Rev. Mod. Phys.}\ }\textbf {\bibinfo {volume}
  {81}},\ \bibinfo {pages} {1229} (\bibinfo {year} {2009})}\BibitemShut
  {NoStop}%
\bibitem [{\citenamefont {Mangles}\ \emph {et~al.}(2004)\citenamefont
  {Mangles}, \citenamefont {Murphy}, \citenamefont {Najmudin}, \citenamefont
  {Thomas}, \citenamefont {Collier}, \citenamefont {Dangor}, \citenamefont
  {Divall}, \citenamefont {Foster}, \citenamefont {Gallacher}, \citenamefont
  {Hooker} \emph {et~al.}}]{ManglesNature2004}%
  \BibitemOpen
  \bibfield  {author} {\bibinfo {author} {\bibfnamefont {S.}~\bibnamefont
  {Mangles}}, \bibinfo {author} {\bibfnamefont {C.}~\bibnamefont {Murphy}},
  \bibinfo {author} {\bibfnamefont {Z.}~\bibnamefont {Najmudin}}, \bibinfo
  {author} {\bibfnamefont {A.}~\bibnamefont {Thomas}}, \bibinfo {author}
  {\bibfnamefont {J.}~\bibnamefont {Collier}}, \bibinfo {author} {\bibfnamefont
  {A.}~\bibnamefont {Dangor}}, \bibinfo {author} {\bibfnamefont
  {E.}~\bibnamefont {Divall}}, \bibinfo {author} {\bibfnamefont
  {P.}~\bibnamefont {Foster}}, \bibinfo {author} {\bibfnamefont
  {J.}~\bibnamefont {Gallacher}}, \bibinfo {author} {\bibfnamefont
  {C.}~\bibnamefont {Hooker}},  \emph {et~al.},\ }\href@noop {} {\bibfield
  {journal} {\bibinfo  {journal} {Nature}\ }\textbf {\bibinfo {volume} {431}},\
  \bibinfo {pages} {535} (\bibinfo {year} {2004})}\BibitemShut {NoStop}%
\bibitem [{\citenamefont {Geddes}\ \emph {et~al.}(2004)\citenamefont {Geddes},
  \citenamefont {Toth}, \citenamefont {Van~Tilborg}, \citenamefont {Esarey},
  \citenamefont {Schroeder}, \citenamefont {Bruhwiler}, \citenamefont {Nieter},
  \citenamefont {Cary},\ and\ \citenamefont {Leemans}}]{GeddesNature2004}%
  \BibitemOpen
  \bibfield  {author} {\bibinfo {author} {\bibfnamefont {C.}~\bibnamefont
  {Geddes}}, \bibinfo {author} {\bibfnamefont {C.}~\bibnamefont {Toth}},
  \bibinfo {author} {\bibfnamefont {J.}~\bibnamefont {Van~Tilborg}}, \bibinfo
  {author} {\bibfnamefont {E.}~\bibnamefont {Esarey}}, \bibinfo {author}
  {\bibfnamefont {C.}~\bibnamefont {Schroeder}}, \bibinfo {author}
  {\bibfnamefont {D.}~\bibnamefont {Bruhwiler}}, \bibinfo {author}
  {\bibfnamefont {C.}~\bibnamefont {Nieter}}, \bibinfo {author} {\bibfnamefont
  {J.}~\bibnamefont {Cary}}, \ and\ \bibinfo {author} {\bibfnamefont
  {W.}~\bibnamefont {Leemans}},\ }\href@noop {} {\bibfield  {journal} {\bibinfo
   {journal} {Nature}\ }\textbf {\bibinfo {volume} {431}},\ \bibinfo {pages}
  {538} (\bibinfo {year} {2004})}\BibitemShut {NoStop}%
\bibitem [{\citenamefont {Faure}\ \emph {et~al.}(2004)\citenamefont {Faure},
  \citenamefont {Glinec}, \citenamefont {Pukhov}, \citenamefont {Kiselev},
  \citenamefont {Gordienko}, \citenamefont {Lefebvre}, \citenamefont
  {Rousseau}, \citenamefont {Burgy},\ and\ \citenamefont
  {Malka}}]{FaureNature2004}%
  \BibitemOpen
  \bibfield  {author} {\bibinfo {author} {\bibfnamefont {J.}~\bibnamefont
  {Faure}}, \bibinfo {author} {\bibfnamefont {Y.}~\bibnamefont {Glinec}},
  \bibinfo {author} {\bibfnamefont {A.}~\bibnamefont {Pukhov}}, \bibinfo
  {author} {\bibfnamefont {S.}~\bibnamefont {Kiselev}}, \bibinfo {author}
  {\bibfnamefont {S.}~\bibnamefont {Gordienko}}, \bibinfo {author}
  {\bibfnamefont {E.}~\bibnamefont {Lefebvre}}, \bibinfo {author}
  {\bibfnamefont {J.-P.}\ \bibnamefont {Rousseau}}, \bibinfo {author}
  {\bibfnamefont {F.}~\bibnamefont {Burgy}}, \ and\ \bibinfo {author}
  {\bibfnamefont {V.}~\bibnamefont {Malka}},\ }\href@noop {} {\bibfield
  {journal} {\bibinfo  {journal} {Nature}\ }\textbf {\bibinfo {volume} {431}},\
  \bibinfo {pages} {541} (\bibinfo {year} {2004})}\BibitemShut {NoStop}%
\bibitem [{\citenamefont {Kim}\ \emph {et~al.}(2013)\citenamefont {Kim},
  \citenamefont {Pae}, \citenamefont {Cha}, \citenamefont {Kim}, \citenamefont
  {Yu}, \citenamefont {Sung}, \citenamefont {Lee}, \citenamefont {Jeong},\ and\
  \citenamefont {Lee}}]{HTKimPRL2013}%
  \BibitemOpen
  \bibfield  {author} {\bibinfo {author} {\bibfnamefont {H.~T.}\ \bibnamefont
  {Kim}}, \bibinfo {author} {\bibfnamefont {K.~H.}\ \bibnamefont {Pae}},
  \bibinfo {author} {\bibfnamefont {H.~J.}\ \bibnamefont {Cha}}, \bibinfo
  {author} {\bibfnamefont {I.~J.}\ \bibnamefont {Kim}}, \bibinfo {author}
  {\bibfnamefont {T.~J.}\ \bibnamefont {Yu}}, \bibinfo {author} {\bibfnamefont
  {J.~H.}\ \bibnamefont {Sung}}, \bibinfo {author} {\bibfnamefont {S.~K.}\
  \bibnamefont {Lee}}, \bibinfo {author} {\bibfnamefont {T.~M.}\ \bibnamefont
  {Jeong}}, \ and\ \bibinfo {author} {\bibfnamefont {J.}~\bibnamefont {Lee}},\
  }\href {\doibase 10.1103/PhysRevLett.111.165002} {\bibfield  {journal}
  {\bibinfo  {journal} {Phys. Rev. Lett.}\ }\textbf {\bibinfo {volume} {111}},\
  \bibinfo {pages} {165002} (\bibinfo {year} {2013})}\BibitemShut {NoStop}%
\bibitem [{\citenamefont {Leemans}\ \emph {et~al.}(2014)\citenamefont
  {Leemans}, \citenamefont {Gonsalves}, \citenamefont {Mao}, \citenamefont
  {Nakamura}, \citenamefont {Benedetti}, \citenamefont {Schroeder},
  \citenamefont {T\'oth}, \citenamefont {Daniels}, \citenamefont
  {Mittelberger}, \citenamefont {Bulanov}, \citenamefont {Vay}, \citenamefont
  {Geddes},\ and\ \citenamefont {Esarey}}]{LeemansPRL2014}%
  \BibitemOpen
  \bibfield  {author} {\bibinfo {author} {\bibfnamefont {W.~P.}\ \bibnamefont
  {Leemans}}, \bibinfo {author} {\bibfnamefont {A.~J.}\ \bibnamefont
  {Gonsalves}}, \bibinfo {author} {\bibfnamefont {H.-S.}\ \bibnamefont {Mao}},
  \bibinfo {author} {\bibfnamefont {K.}~\bibnamefont {Nakamura}}, \bibinfo
  {author} {\bibfnamefont {C.}~\bibnamefont {Benedetti}}, \bibinfo {author}
  {\bibfnamefont {C.~B.}\ \bibnamefont {Schroeder}}, \bibinfo {author}
  {\bibfnamefont {C.}~\bibnamefont {T\'oth}}, \bibinfo {author} {\bibfnamefont
  {J.}~\bibnamefont {Daniels}}, \bibinfo {author} {\bibfnamefont {D.~E.}\
  \bibnamefont {Mittelberger}}, \bibinfo {author} {\bibfnamefont {S.~S.}\
  \bibnamefont {Bulanov}}, \bibinfo {author} {\bibfnamefont {J.-L.}\
  \bibnamefont {Vay}}, \bibinfo {author} {\bibfnamefont {C.~G.~R.}\
  \bibnamefont {Geddes}}, \ and\ \bibinfo {author} {\bibfnamefont
  {E.}~\bibnamefont {Esarey}},\ }\href {\doibase
  10.1103/PhysRevLett.113.245002} {\bibfield  {journal} {\bibinfo  {journal}
  {Phys. Rev. Lett.}\ }\textbf {\bibinfo {volume} {113}},\ \bibinfo {pages}
  {245002} (\bibinfo {year} {2014})}\BibitemShut {NoStop}%
\bibitem [{\citenamefont {Homma}\ \emph {et~al.}(2016)\citenamefont {Homma},
  \citenamefont {Tesileanu}, \citenamefont {D'Alessi}, \citenamefont {Hasebe},
  \citenamefont {Ilderton}, \citenamefont {Moritaka}, \citenamefont {Nakamiya},
  \citenamefont {Seto},\ and\ \citenamefont {Utsunomiya}}]{KHommaRRP2016}%
  \BibitemOpen
  \bibfield  {author} {\bibinfo {author} {\bibfnamefont {K.}~\bibnamefont
  {Homma}}, \bibinfo {author} {\bibfnamefont {O.}~\bibnamefont {Tesileanu}},
  \bibinfo {author} {\bibfnamefont {L.}~\bibnamefont {D'Alessi}}, \bibinfo
  {author} {\bibfnamefont {T.}~\bibnamefont {Hasebe}}, \bibinfo {author}
  {\bibfnamefont {A.}~\bibnamefont {Ilderton}}, \bibinfo {author}
  {\bibfnamefont {T.}~\bibnamefont {Moritaka}}, \bibinfo {author}
  {\bibfnamefont {Y.}~\bibnamefont {Nakamiya}}, \bibinfo {author}
  {\bibfnamefont {K.}~\bibnamefont {Seto}}, \ and\ \bibinfo {author}
  {\bibfnamefont {H.}~\bibnamefont {Utsunomiya}},\ }\href@noop {} {\bibfield
  {journal} {\bibinfo  {journal} {Romanian Reports in Physics}\ }\textbf
  {\bibinfo {volume} {68}},\ \bibinfo {pages} {S233} (\bibinfo {year}
  {2016})}\BibitemShut {NoStop}%
\bibitem [{\citenamefont {Shen}\ \emph {et~al.}(2012)\citenamefont {Shen},
  \citenamefont {Wu}, \citenamefont {Dong}, \citenamefont {Zhu}, \citenamefont
  {Gu}, \citenamefont {Ji}, \citenamefont {Jiao}, \citenamefont {Teng},
  \citenamefont {Hong}, \citenamefont {Zhao}, \citenamefont {Cao},
  \citenamefont {Wang},\ and\ \citenamefont {Yu}}]{BFShenPOP2012}%
  \BibitemOpen
  \bibfield  {author} {\bibinfo {author} {\bibfnamefont {B.}~\bibnamefont
  {Shen}}, \bibinfo {author} {\bibfnamefont {Y.}~\bibnamefont {Wu}}, \bibinfo
  {author} {\bibfnamefont {K.}~\bibnamefont {Dong}}, \bibinfo {author}
  {\bibfnamefont {B.}~\bibnamefont {Zhu}}, \bibinfo {author} {\bibfnamefont
  {Y.}~\bibnamefont {Gu}}, \bibinfo {author} {\bibfnamefont {L.}~\bibnamefont
  {Ji}}, \bibinfo {author} {\bibfnamefont {C.}~\bibnamefont {Jiao}}, \bibinfo
  {author} {\bibfnamefont {J.}~\bibnamefont {Teng}}, \bibinfo {author}
  {\bibfnamefont {W.}~\bibnamefont {Hong}}, \bibinfo {author} {\bibfnamefont
  {Z.}~\bibnamefont {Zhao}}, \bibinfo {author} {\bibfnamefont {L.}~\bibnamefont
  {Cao}}, \bibinfo {author} {\bibfnamefont {X.}~\bibnamefont {Wang}}, \ and\
  \bibinfo {author} {\bibfnamefont {M.~Y.}\ \bibnamefont {Yu}},\ }\href
  {\doibase http://dx.doi.org/10.1063/1.3694679} {\bibfield  {journal}
  {\bibinfo  {journal} {Phys. Plasmas}\ }\textbf {\bibinfo {volume} {19}},\
  \bibinfo {eid} {033106} (\bibinfo {year} {2012})}\BibitemShut {NoStop}%
\bibitem [{\citenamefont {Sprangle}\ \emph {et~al.}(1987)\citenamefont
  {Sprangle}, \citenamefont {Tang},\ and\ \citenamefont
  {Esarey}}]{SprangleIEEE1987}%
  \BibitemOpen
  \bibfield  {author} {\bibinfo {author} {\bibfnamefont {P.}~\bibnamefont
  {Sprangle}}, \bibinfo {author} {\bibfnamefont {C.-M.}\ \bibnamefont {Tang}},
  \ and\ \bibinfo {author} {\bibfnamefont {E.}~\bibnamefont {Esarey}},\
  }\href@noop {} {\bibfield  {journal} {\bibinfo  {journal} {IEEE transactions
  on plasma science}\ }\textbf {\bibinfo {volume} {15}},\ \bibinfo {pages}
  {145} (\bibinfo {year} {1987})}\BibitemShut {NoStop}%
\bibitem [{\citenamefont {Sun}\ \emph {et~al.}(1987)\citenamefont {Sun},
  \citenamefont {Ott}, \citenamefont {Lee},\ and\ \citenamefont
  {Guzdar}}]{GZSunPOF1987}%
  \BibitemOpen
  \bibfield  {author} {\bibinfo {author} {\bibfnamefont {G.~Z.}\ \bibnamefont
  {Sun}}, \bibinfo {author} {\bibfnamefont {E.}~\bibnamefont {Ott}}, \bibinfo
  {author} {\bibfnamefont {Y.~C.}\ \bibnamefont {Lee}}, \ and\ \bibinfo
  {author} {\bibfnamefont {P.}~\bibnamefont {Guzdar}},\ }\href {\doibase
  http://dx.doi.org/10.1063/1.866349} {\bibfield  {journal} {\bibinfo
  {journal} {Phys. Fluids}\ }\textbf {\bibinfo {volume} {30}},\ \bibinfo
  {pages} {526} (\bibinfo {year} {1987})}\BibitemShut {NoStop}%
\bibitem [{\citenamefont {Tzeng}\ and\ \citenamefont
  {Mori}(1998)}]{KCTzengPRL1998}%
  \BibitemOpen
  \bibfield  {author} {\bibinfo {author} {\bibfnamefont {K.-C.}\ \bibnamefont
  {Tzeng}}\ and\ \bibinfo {author} {\bibfnamefont {W.~B.}\ \bibnamefont
  {Mori}},\ }\href {\doibase 10.1103/PhysRevLett.81.104} {\bibfield  {journal}
  {\bibinfo  {journal} {Phys. Rev. Lett.}\ }\textbf {\bibinfo {volume} {81}},\
  \bibinfo {pages} {104} (\bibinfo {year} {1998})}\BibitemShut {NoStop}%
\bibitem [{\citenamefont {Zeng}\ \emph {et~al.}(2014)\citenamefont {Zeng},
  \citenamefont {Chen}, \citenamefont {Sheng}, \citenamefont {Mori},\ and\
  \citenamefont {Zhang}}]{MZengPOP2014}%
  \BibitemOpen
  \bibfield  {author} {\bibinfo {author} {\bibfnamefont {M.}~\bibnamefont
  {Zeng}}, \bibinfo {author} {\bibfnamefont {M.}~\bibnamefont {Chen}}, \bibinfo
  {author} {\bibfnamefont {Z.-M.}\ \bibnamefont {Sheng}}, \bibinfo {author}
  {\bibfnamefont {W.~B.}\ \bibnamefont {Mori}}, \ and\ \bibinfo {author}
  {\bibfnamefont {J.}~\bibnamefont {Zhang}},\ }\href {\doibase
  http://dx.doi.org/10.1063/1.4868404} {\bibfield  {journal} {\bibinfo
  {journal} {Phys. Plasmas}\ }\textbf {\bibinfo {volume} {21}},\ \bibinfo {eid}
  {030701} (\bibinfo {year} {2014})}\BibitemShut {NoStop}%
\bibitem [{\citenamefont {Zeng}\ \emph
  {et~al.}(2016{\natexlab{a}})\citenamefont {Zeng}, \citenamefont {Chen},\ and\
  \citenamefont {Sheng}}]{MZengJPCS2016}%
  \BibitemOpen
  \bibfield  {author} {\bibinfo {author} {\bibfnamefont {M.}~\bibnamefont
  {Zeng}}, \bibinfo {author} {\bibfnamefont {M.}~\bibnamefont {Chen}}, \ and\
  \bibinfo {author} {\bibfnamefont {Z.-M.}\ \bibnamefont {Sheng}},\ }\href
  {http://stacks.iop.org/1742-6596/688/i=1/a=012130} {\bibfield  {journal}
  {\bibinfo  {journal} {Journal of Physics: Conference Series}\ }\textbf
  {\bibinfo {volume} {688}},\ \bibinfo {pages} {012130} (\bibinfo {year}
  {2016}{\natexlab{a}})}\BibitemShut {NoStop}%
\bibitem [{\citenamefont {Chen}\ \emph {et~al.}(2006)\citenamefont {Chen},
  \citenamefont {Sheng}, \citenamefont {Ma},\ and\ \citenamefont
  {Zhang}}]{MinChenJAP2006}%
  \BibitemOpen
  \bibfield  {author} {\bibinfo {author} {\bibfnamefont {M.}~\bibnamefont
  {Chen}}, \bibinfo {author} {\bibfnamefont {Z.-M.}\ \bibnamefont {Sheng}},
  \bibinfo {author} {\bibfnamefont {Y.-Y.}\ \bibnamefont {Ma}}, \ and\ \bibinfo
  {author} {\bibfnamefont {J.}~\bibnamefont {Zhang}},\ }\href {\doibase
  10.1063/1.2179194} {\bibfield  {journal} {\bibinfo  {journal} {J. Appl.
  Phys.}\ }\textbf {\bibinfo {volume} {99}},\ \bibinfo {pages} {056109}
  (\bibinfo {year} {2006})}\BibitemShut {NoStop}%
\bibitem [{\citenamefont {McGuffey}\ \emph {et~al.}(2010)\citenamefont
  {McGuffey}, \citenamefont {Thomas}, \citenamefont {Schumaker}, \citenamefont
  {Matsuoka}, \citenamefont {Chvykov}, \citenamefont {Dollar}, \citenamefont
  {Kalintchenko}, \citenamefont {Yanovsky}, \citenamefont {Maksimchuk},
  \citenamefont {Krushelnick}, \citenamefont {Bychenkov}, \citenamefont
  {Glazyrin},\ and\ \citenamefont {Karpeev}}]{McGuffeyPRL2010}%
  \BibitemOpen
  \bibfield  {author} {\bibinfo {author} {\bibfnamefont {C.}~\bibnamefont
  {McGuffey}}, \bibinfo {author} {\bibfnamefont {A.~G.~R.}\ \bibnamefont
  {Thomas}}, \bibinfo {author} {\bibfnamefont {W.}~\bibnamefont {Schumaker}},
  \bibinfo {author} {\bibfnamefont {T.}~\bibnamefont {Matsuoka}}, \bibinfo
  {author} {\bibfnamefont {V.}~\bibnamefont {Chvykov}}, \bibinfo {author}
  {\bibfnamefont {F.~J.}\ \bibnamefont {Dollar}}, \bibinfo {author}
  {\bibfnamefont {G.}~\bibnamefont {Kalintchenko}}, \bibinfo {author}
  {\bibfnamefont {V.}~\bibnamefont {Yanovsky}}, \bibinfo {author}
  {\bibfnamefont {A.}~\bibnamefont {Maksimchuk}}, \bibinfo {author}
  {\bibfnamefont {K.}~\bibnamefont {Krushelnick}}, \bibinfo {author}
  {\bibfnamefont {V.~Y.}\ \bibnamefont {Bychenkov}}, \bibinfo {author}
  {\bibfnamefont {I.~V.}\ \bibnamefont {Glazyrin}}, \ and\ \bibinfo {author}
  {\bibfnamefont {A.~V.}\ \bibnamefont {Karpeev}},\ }\href {\doibase
  10.1103/PhysRevLett.104.025004} {\bibfield  {journal} {\bibinfo  {journal}
  {Phys. Rev. Lett.}\ }\textbf {\bibinfo {volume} {104}},\ \bibinfo {pages}
  {025004} (\bibinfo {year} {2010})}\BibitemShut {NoStop}%
\bibitem [{\citenamefont {Pak}\ \emph {et~al.}(2010)\citenamefont {Pak},
  \citenamefont {Marsh}, \citenamefont {Martins}, \citenamefont {Lu},
  \citenamefont {Mori},\ and\ \citenamefont {Joshi}}]{PakPRL2010}%
  \BibitemOpen
  \bibfield  {author} {\bibinfo {author} {\bibfnamefont {A.}~\bibnamefont
  {Pak}}, \bibinfo {author} {\bibfnamefont {K.~A.}\ \bibnamefont {Marsh}},
  \bibinfo {author} {\bibfnamefont {S.~F.}\ \bibnamefont {Martins}}, \bibinfo
  {author} {\bibfnamefont {W.}~\bibnamefont {Lu}}, \bibinfo {author}
  {\bibfnamefont {W.~B.}\ \bibnamefont {Mori}}, \ and\ \bibinfo {author}
  {\bibfnamefont {C.}~\bibnamefont {Joshi}},\ }\href {\doibase
  10.1103/PhysRevLett.104.025003} {\bibfield  {journal} {\bibinfo  {journal}
  {Phys. Rev. Lett.}\ }\textbf {\bibinfo {volume} {104}},\ \bibinfo {pages}
  {025003} (\bibinfo {year} {2010})}\BibitemShut {NoStop}%
\bibitem [{\citenamefont {Clayton}\ \emph {et~al.}(2010)\citenamefont
  {Clayton}, \citenamefont {Ralph}, \citenamefont {Albert}, \citenamefont
  {Fonseca}, \citenamefont {Glenzer}, \citenamefont {Joshi}, \citenamefont
  {Lu}, \citenamefont {Marsh}, \citenamefont {Martins}, \citenamefont {Mori},
  \citenamefont {Pak}, \citenamefont {Tsung}, \citenamefont {Pollock},
  \citenamefont {Ross}, \citenamefont {Silva},\ and\ \citenamefont
  {Froula}}]{ClaytonPRL2010}%
  \BibitemOpen
  \bibfield  {author} {\bibinfo {author} {\bibfnamefont {C.~E.}\ \bibnamefont
  {Clayton}}, \bibinfo {author} {\bibfnamefont {J.~E.}\ \bibnamefont {Ralph}},
  \bibinfo {author} {\bibfnamefont {F.}~\bibnamefont {Albert}}, \bibinfo
  {author} {\bibfnamefont {R.~A.}\ \bibnamefont {Fonseca}}, \bibinfo {author}
  {\bibfnamefont {S.~H.}\ \bibnamefont {Glenzer}}, \bibinfo {author}
  {\bibfnamefont {C.}~\bibnamefont {Joshi}}, \bibinfo {author} {\bibfnamefont
  {W.}~\bibnamefont {Lu}}, \bibinfo {author} {\bibfnamefont {K.~A.}\
  \bibnamefont {Marsh}}, \bibinfo {author} {\bibfnamefont {S.~F.}\ \bibnamefont
  {Martins}}, \bibinfo {author} {\bibfnamefont {W.~B.}\ \bibnamefont {Mori}},
  \bibinfo {author} {\bibfnamefont {A.}~\bibnamefont {Pak}}, \bibinfo {author}
  {\bibfnamefont {F.~S.}\ \bibnamefont {Tsung}}, \bibinfo {author}
  {\bibfnamefont {B.~B.}\ \bibnamefont {Pollock}}, \bibinfo {author}
  {\bibfnamefont {J.~S.}\ \bibnamefont {Ross}}, \bibinfo {author}
  {\bibfnamefont {L.~O.}\ \bibnamefont {Silva}}, \ and\ \bibinfo {author}
  {\bibfnamefont {D.~H.}\ \bibnamefont {Froula}},\ }\href {\doibase
  10.1103/PhysRevLett.105.105003} {\bibfield  {journal} {\bibinfo  {journal}
  {Phys. Rev. Lett.}\ }\textbf {\bibinfo {volume} {105}},\ \bibinfo {pages}
  {105003} (\bibinfo {year} {2010})}\BibitemShut {NoStop}%
\bibitem [{\citenamefont {Liu}\ \emph {et~al.}(2011)\citenamefont {Liu},
  \citenamefont {Xia}, \citenamefont {Wang}, \citenamefont {Lu}, \citenamefont
  {Wang}, \citenamefont {Deng}, \citenamefont {Li}, \citenamefont {Zhang},
  \citenamefont {Liang}, \citenamefont {Leng}, \citenamefont {Lu},
  \citenamefont {Wang}, \citenamefont {Wang}, \citenamefont {Nakajima},
  \citenamefont {Li},\ and\ \citenamefont {Xu}}]{JSLiuPRL2011}%
  \BibitemOpen
  \bibfield  {author} {\bibinfo {author} {\bibfnamefont {J.~S.}\ \bibnamefont
  {Liu}}, \bibinfo {author} {\bibfnamefont {C.~Q.}\ \bibnamefont {Xia}},
  \bibinfo {author} {\bibfnamefont {W.~T.}\ \bibnamefont {Wang}}, \bibinfo
  {author} {\bibfnamefont {H.~Y.}\ \bibnamefont {Lu}}, \bibinfo {author}
  {\bibfnamefont {C.}~\bibnamefont {Wang}}, \bibinfo {author} {\bibfnamefont
  {A.~H.}\ \bibnamefont {Deng}}, \bibinfo {author} {\bibfnamefont {W.~T.}\
  \bibnamefont {Li}}, \bibinfo {author} {\bibfnamefont {H.}~\bibnamefont
  {Zhang}}, \bibinfo {author} {\bibfnamefont {X.~Y.}\ \bibnamefont {Liang}},
  \bibinfo {author} {\bibfnamefont {Y.~X.}\ \bibnamefont {Leng}}, \bibinfo
  {author} {\bibfnamefont {X.~M.}\ \bibnamefont {Lu}}, \bibinfo {author}
  {\bibfnamefont {C.}~\bibnamefont {Wang}}, \bibinfo {author} {\bibfnamefont
  {J.~Z.}\ \bibnamefont {Wang}}, \bibinfo {author} {\bibfnamefont
  {K.}~\bibnamefont {Nakajima}}, \bibinfo {author} {\bibfnamefont {R.~X.}\
  \bibnamefont {Li}}, \ and\ \bibinfo {author} {\bibfnamefont {Z.~Z.}\
  \bibnamefont {Xu}},\ }\href {\doibase 10.1103/PhysRevLett.107.035001}
  {\bibfield  {journal} {\bibinfo  {journal} {Phys. Rev. Lett.}\ }\textbf
  {\bibinfo {volume} {107}},\ \bibinfo {pages} {035001} (\bibinfo {year}
  {2011})}\BibitemShut {NoStop}%
\bibitem [{\citenamefont {Pollock}\ \emph {et~al.}(2011)\citenamefont
  {Pollock}, \citenamefont {Clayton}, \citenamefont {Ralph}, \citenamefont
  {Albert}, \citenamefont {Davidson}, \citenamefont {Divol}, \citenamefont
  {Filip}, \citenamefont {Glenzer}, \citenamefont {Herpoldt}, \citenamefont
  {Lu}, \citenamefont {Marsh}, \citenamefont {Meinecke}, \citenamefont {Mori},
  \citenamefont {Pak}, \citenamefont {Rensink}, \citenamefont {Ross},
  \citenamefont {Shaw}, \citenamefont {Tynan}, \citenamefont {Joshi},\ and\
  \citenamefont {Froula}}]{PollockPRL2011}%
  \BibitemOpen
  \bibfield  {author} {\bibinfo {author} {\bibfnamefont {B.~B.}\ \bibnamefont
  {Pollock}}, \bibinfo {author} {\bibfnamefont {C.~E.}\ \bibnamefont
  {Clayton}}, \bibinfo {author} {\bibfnamefont {J.~E.}\ \bibnamefont {Ralph}},
  \bibinfo {author} {\bibfnamefont {F.}~\bibnamefont {Albert}}, \bibinfo
  {author} {\bibfnamefont {A.}~\bibnamefont {Davidson}}, \bibinfo {author}
  {\bibfnamefont {L.}~\bibnamefont {Divol}}, \bibinfo {author} {\bibfnamefont
  {C.}~\bibnamefont {Filip}}, \bibinfo {author} {\bibfnamefont {S.~H.}\
  \bibnamefont {Glenzer}}, \bibinfo {author} {\bibfnamefont {K.}~\bibnamefont
  {Herpoldt}}, \bibinfo {author} {\bibfnamefont {W.}~\bibnamefont {Lu}},
  \bibinfo {author} {\bibfnamefont {K.~A.}\ \bibnamefont {Marsh}}, \bibinfo
  {author} {\bibfnamefont {J.}~\bibnamefont {Meinecke}}, \bibinfo {author}
  {\bibfnamefont {W.~B.}\ \bibnamefont {Mori}}, \bibinfo {author}
  {\bibfnamefont {A.}~\bibnamefont {Pak}}, \bibinfo {author} {\bibfnamefont
  {T.~C.}\ \bibnamefont {Rensink}}, \bibinfo {author} {\bibfnamefont {J.~S.}\
  \bibnamefont {Ross}}, \bibinfo {author} {\bibfnamefont {J.}~\bibnamefont
  {Shaw}}, \bibinfo {author} {\bibfnamefont {G.~R.}\ \bibnamefont {Tynan}},
  \bibinfo {author} {\bibfnamefont {C.}~\bibnamefont {Joshi}}, \ and\ \bibinfo
  {author} {\bibfnamefont {D.~H.}\ \bibnamefont {Froula}},\ }\href {\doibase
  10.1103/PhysRevLett.107.045001} {\bibfield  {journal} {\bibinfo  {journal}
  {Phys. Rev. Lett.}\ }\textbf {\bibinfo {volume} {107}},\ \bibinfo {pages}
  {045001} (\bibinfo {year} {2011})}\BibitemShut {NoStop}%
\bibitem [{\citenamefont {Zeng}\ \emph {et~al.}(2012)\citenamefont {Zeng},
  \citenamefont {Hafz}, \citenamefont {Nakajima}, \citenamefont {Chen},
  \citenamefont {Lu}, \citenamefont {Mori}, \citenamefont {Sheng},\ and\
  \citenamefont {Zhang}}]{MZengJPP2012}%
  \BibitemOpen
  \bibfield  {author} {\bibinfo {author} {\bibfnamefont {M.}~\bibnamefont
  {Zeng}}, \bibinfo {author} {\bibfnamefont {N.~A.~M.}\ \bibnamefont {Hafz}},
  \bibinfo {author} {\bibfnamefont {K.}~\bibnamefont {Nakajima}}, \bibinfo
  {author} {\bibfnamefont {L.-M.}\ \bibnamefont {Chen}}, \bibinfo {author}
  {\bibfnamefont {W.}~\bibnamefont {Lu}}, \bibinfo {author} {\bibfnamefont
  {W.~B.}\ \bibnamefont {Mori}}, \bibinfo {author} {\bibfnamefont {Z.-M.}\
  \bibnamefont {Sheng}}, \ and\ \bibinfo {author} {\bibfnamefont
  {J.}~\bibnamefont {Zhang}},\ }\href {\doibase 10.1017/S0022377812000098}
  {\bibfield  {journal} {\bibinfo  {journal} {J. Plasma Phys.}\ }\textbf
  {\bibinfo {volume} {78}},\ \bibinfo {pages} {363} (\bibinfo {year}
  {2012})}\BibitemShut {NoStop}%
\bibitem [{\citenamefont {Chen}\ \emph {et~al.}(2013)\citenamefont {Chen},
  \citenamefont {Cormier-Michel}, \citenamefont {Geddes}, \citenamefont
  {Bruhwiler}, \citenamefont {Yu}, \citenamefont {Esarey}, \citenamefont
  {Schroeder},\ and\ \citenamefont {Leemans}}]{MChenJCP2013}%
  \BibitemOpen
  \bibfield  {author} {\bibinfo {author} {\bibfnamefont {M.}~\bibnamefont
  {Chen}}, \bibinfo {author} {\bibfnamefont {E.}~\bibnamefont
  {Cormier-Michel}}, \bibinfo {author} {\bibfnamefont {C.}~\bibnamefont
  {Geddes}}, \bibinfo {author} {\bibfnamefont {D.}~\bibnamefont {Bruhwiler}},
  \bibinfo {author} {\bibfnamefont {L.}~\bibnamefont {Yu}}, \bibinfo {author}
  {\bibfnamefont {E.}~\bibnamefont {Esarey}}, \bibinfo {author} {\bibfnamefont
  {C.}~\bibnamefont {Schroeder}}, \ and\ \bibinfo {author} {\bibfnamefont
  {W.}~\bibnamefont {Leemans}},\ }\href {\doibase
  http://dx.doi.org/10.1016/j.jcp.2012.11.029} {\bibfield  {journal} {\bibinfo
  {journal} {J. Comput. Phys.}\ }\textbf {\bibinfo {volume} {236}},\ \bibinfo
  {pages} {220 } (\bibinfo {year} {2013})}\BibitemShut {NoStop}%
\bibitem [{\citenamefont {Zeng}\ \emph {et~al.}(2015)\citenamefont {Zeng},
  \citenamefont {Chen}, \citenamefont {Yu}, \citenamefont {Mori}, \citenamefont
  {Sheng}, \citenamefont {Hidding}, \citenamefont {Jaroszynski},\ and\
  \citenamefont {Zhang}}]{MZengPRL2015}%
  \BibitemOpen
  \bibfield  {author} {\bibinfo {author} {\bibfnamefont {M.}~\bibnamefont
  {Zeng}}, \bibinfo {author} {\bibfnamefont {M.}~\bibnamefont {Chen}}, \bibinfo
  {author} {\bibfnamefont {L.~L.}\ \bibnamefont {Yu}}, \bibinfo {author}
  {\bibfnamefont {W.~B.}\ \bibnamefont {Mori}}, \bibinfo {author}
  {\bibfnamefont {Z.~M.}\ \bibnamefont {Sheng}}, \bibinfo {author}
  {\bibfnamefont {B.}~\bibnamefont {Hidding}}, \bibinfo {author} {\bibfnamefont
  {D.~A.}\ \bibnamefont {Jaroszynski}}, \ and\ \bibinfo {author} {\bibfnamefont
  {J.}~\bibnamefont {Zhang}},\ }\href {\doibase 10.1103/PhysRevLett.114.084801}
  {\bibfield  {journal} {\bibinfo  {journal} {Phys. Rev. Lett.}\ }\textbf
  {\bibinfo {volume} {114}},\ \bibinfo {pages} {084801} (\bibinfo {year}
  {2015})}\BibitemShut {NoStop}%
\bibitem [{\citenamefont {Zeng}\ \emph
  {et~al.}(2016{\natexlab{b}})\citenamefont {Zeng}, \citenamefont {Luo},
  \citenamefont {Chen}, \citenamefont {Mori}, \citenamefont {Sheng},\ and\
  \citenamefont {Hidding}}]{MZengPOP2016}%
  \BibitemOpen
  \bibfield  {author} {\bibinfo {author} {\bibfnamefont {M.}~\bibnamefont
  {Zeng}}, \bibinfo {author} {\bibfnamefont {J.}~\bibnamefont {Luo}}, \bibinfo
  {author} {\bibfnamefont {M.}~\bibnamefont {Chen}}, \bibinfo {author}
  {\bibfnamefont {W.~B.}\ \bibnamefont {Mori}}, \bibinfo {author}
  {\bibfnamefont {Z.-M.}\ \bibnamefont {Sheng}}, \ and\ \bibinfo {author}
  {\bibfnamefont {B.}~\bibnamefont {Hidding}},\ }\href {\doibase
  http://dx.doi.org/10.1063/1.4953895} {\bibfield  {journal} {\bibinfo
  {journal} {Phys. Plasmas}\ }\textbf {\bibinfo {volume} {23}},\ \bibinfo {eid}
  {063113} (\bibinfo {year} {2016}{\natexlab{b}})}\BibitemShut {NoStop}%
\bibitem [{\citenamefont {Arber}\ \emph {et~al.}(2015)\citenamefont {Arber},
  \citenamefont {Bennett}, \citenamefont {Brady}, \citenamefont
  {Lawrence-Douglas}, \citenamefont {Ramsay}, \citenamefont {Sircombe},
  \citenamefont {Gillies}, \citenamefont {Evans}, \citenamefont {Schmitz},
  \citenamefont {Bell},\ and\ \citenamefont {Ridgers}}]{EPOCH}%
  \BibitemOpen
  \bibfield  {author} {\bibinfo {author} {\bibfnamefont {T.~D.}\ \bibnamefont
  {Arber}}, \bibinfo {author} {\bibfnamefont {K.}~\bibnamefont {Bennett}},
  \bibinfo {author} {\bibfnamefont {C.~S.}\ \bibnamefont {Brady}}, \bibinfo
  {author} {\bibfnamefont {A.}~\bibnamefont {Lawrence-Douglas}}, \bibinfo
  {author} {\bibfnamefont {M.~G.}\ \bibnamefont {Ramsay}}, \bibinfo {author}
  {\bibfnamefont {N.~J.}\ \bibnamefont {Sircombe}}, \bibinfo {author}
  {\bibfnamefont {P.}~\bibnamefont {Gillies}}, \bibinfo {author} {\bibfnamefont
  {R.~G.}\ \bibnamefont {Evans}}, \bibinfo {author} {\bibfnamefont
  {H.}~\bibnamefont {Schmitz}}, \bibinfo {author} {\bibfnamefont {A.~R.}\
  \bibnamefont {Bell}}, \ and\ \bibinfo {author} {\bibfnamefont {C.~P.}\
  \bibnamefont {Ridgers}},\ }\href@noop {} {\bibfield  {journal} {\bibinfo
  {journal} {Plasma Physics and Controlled Fusion}\ }\textbf {\bibinfo {volume}
  {57}},\ \bibinfo {pages} {1} (\bibinfo {year} {2015})}\BibitemShut {NoStop}%
\end{thebibliography}%
\end{document}